# Generative Retrieval under Real-Time Supply Constraints in Local Commerce

Shaopeng Liang*, Biao Zhang*, Hongzhen Ye*, Zeyang Hou*, Ma Jiang*
*Rajax Network Technology (Taobao Shangou of Alibaba)*
Shanghai, China
{shaopengliang.lsp, shangxin.zb, houzeyang.hzy, majiang.mj}@alibaba-inc.com
yehongzhen1999@163.com

Jintao Tu, yunxi.hz, Zhijia Fang, Jianqiang Cao, Junxue Rao, Jia Jia
*Rajax Network Technology (Taobao Shangou of Alibaba)*
Shanghai, China
{tujintao.tjt, yunxi.hz, jianqiang.caojq, junxue.rao, jj229618}@alibaba-inc.com
zhijia.fzj@koubei.com

* These authors contributed equally to this work.

## Abstract

Generative retrieval predicts semantic identifiers for candidate entities, offering a learned alternative to conventional candidate lookup. In local retail and food delivery, those candidates must also be serviceable for a particular user at a particular time. Location, weather, and traffic reshape available merchant and product supply, so a catalog-valid identifier may be infeasible for the current request. Post-generation filtering removes such candidates after beam capacity and inference computation have already been spent. We study generative retrieval over an explicitly materialized, request-time supply snapshot. This setting puts constraint preparation on the online serving path, where short output sequences provide little opportunity to amortize compilation and transfer. We present a pipeline integrated into RTP-LLM that compiles eligible residual-quantization identifiers into a compressed sparse row prefix index, overlaps CPU compilation with model prefill, and shares the uploaded index across beams. Legal expansions retain original-model log probabilities, and constraint states follow selected beam parents. We formalize snapshot-relative identifier validity and distinguish it from entity-level eligibility and freshness. The examined implementation generates three-token merchant identifiers. We analyze historical evidence from T+1 daily-refresh catalog constraints and subsequent request-scoped serving evaluations. In a matched-beam daily-constraint case, merchant filtering falls from 170 of 299 candidates to 29 of 299, while residual distance filtering motivates request-specific supply constraints. Later records report approximately 3 ms for C++ sorting and 1 ms for CSR construction. A different-host historical comparison with 30,000 supplied entries reports a 0.6 ms service-side latency increment, but broader client measurements expose additional request costs. Captured Base64 requests from three geographic settings further characterize behavior across supply sizes and effective output widths. These observations establish a practical implementation and identify integration bottlenecks; they do not yet establish compute savings at matched retrieval quality. The resulting framework connects changing supply, constrained search, and the cost of producing useful eligible results.

## 1 Introduction

Generative retrieval is a developing direction in information retrieval and recommendation: a model retrieves an entity by generating its identifier. Work on autoregressive entity retrieval and semantic identifiers has established alternatives to conventional candidate lookup [1, 2]. This direction offers a way to learn candidate generation as a sequence prediction problem. In our setting, residual quantization produces a short semantic identifier (SID), and the retrieval model generates SID tokens using beam search. The opportunity is to apply this learned retrieval mechanism to merchant and product discovery while respecting the supply that is actually available.

A local-commerce retrieval system must answer two questions at once: what is relevant to the user, and what can actually be supplied to that user now? The second question is especially demanding in local retail and food delivery. The user's location determines which merchants can serve the request. Weather and traffic can change delivery feasibility, so the eligible supply may change even when the user, merchant, and product catalog remain unchanged. A retrieval decision that was valid for a nearby user or an earlier request may no longer be valid for the current one. A model trained over a global candidate space does not automatically track these request-time changes.

This mismatch creates two distinct errors. A model may generate an identifier that does not correspond to a catalog entity. It may also generate a valid identifier whose merchant or product cannot serve the current request. We call the latter a **supply hallucination**: the output refers to real supply, but incorrectly treats it as available in the request's context. This term concerns retrieval eligibility, not the factual accuracy of natural-language text. Preventing catalog-invalid identifiers alone does not prevent supply hallucinations.

Post-generation filtering can remove unavailable candidates before they reach the user. However, it cannot recover the beam slots already spent exploring them. If many high-scoring paths are ineligible, the final candidate set may be too small, prompting wider decoding, additional attempts, or fallback retrieval. These mechanisms consume resources without necessarily improving the set of usable results. The objective is therefore to direct decoding toward currently eligible supply and reduce the computation required to obtain a useful candidate set.

Request-scoped constraints introduce a systems challenge of their own. The eligible set can differ across consecutive requests and may contain tens of thousands of SIDs. Serializing this set, compiling it into a prefix structure, transferring it to an accelerator, and consulting it during beam search all add cost. For a short SID, there are only a few decoding steps over which to amortize preparation. A fast GPU membership test is insufficient if request preparation dominates latency. Moreover, filtering must preserve the intended ranking objective, and the constraint state must remain aligned with each beam as parents are selected, duplicated, or discarded.

We address these issues with a request-scoped constraint pipeline integrated into RTP-LLM. The upstream supply system materializes the eligible SID set. A background CPU task parses, sorts, and compiles that set while the model processes the prompt. Before the first constrained sampling step, the engine waits for the index and uploads it to the GPU. Beams share the immutable index but maintain separate prefix states. Every decoding step masks illegal continuations, compares legal candidates using original-model log probabilities, and propagates the selected parent mapping to both model and constraint state.

Our focus is **generative retrieval over dynamically changing merchant and product supply**. Constrained decoding is the mechanism that connects this retrieval objective to request-time feasibility. The system treats the changing supply set across request preparation, constrained search, and output resolution, rather than addressing a mask kernel in isolation. CSR storage and probability-preserving masking are established components on which this design builds. This paper makes three concrete contributions:

1. A formulation of supply-constrained generative retrieval that distinguishes catalog validity, request eligibility, and freshness, and connects these properties to useful retrieval cost.

2. An implementation of request-time constraint compilation and beam execution, including prefill overlap, shared accelerator indexes, and parent-aligned constraint states.

3. An empirical account of daily-constraint retrieval cases, request-scoped serving measurements, and captured-request replay, accompanied by validity and score invariants and a cost decomposition that explains the observed bottlenecks.

The business problem includes both merchant and product supply. The implementation examined here uses three-token SIDs for merchant retrieval. The same finite-set formulation can express product eligibility when the corresponding mappings and eligible sets are available; the merchant implementation does not by itself establish a completed product-level deployment.

# 2 Problem Definition

## 2.1 Supply eligibility at request time

Let x denote a request, including its user context and location, and let τ denote the time of the supply snapshot used to process it. Let $E_{x,\tau}$ be the set of entities that the upstream supply service considers eligible for this request. Weather and traffic affect this set through supply and delivery decisions. The decoder consumes the resulting eligible set; it does not independently infer delivery feasibility from weather or traffic inputs.

Let φ map an entity to a tokenized SID of length L over vocabulary V. The request constraint is

$$S_{x,\tau} = \{ \varphi(e) : e \in E_{x,\tau} \} \subseteq V^L. \quad (1)$$

The desired retrieval objective is to find high-scoring identifiers in $S_{x,\tau}$ under the model distribution:

$$Y^* = \mathrm{TopK}_{y \in S_{x,\tau}} \log p_\theta(y \mid x). \quad (2)$$

Beam search approximates this objective. A finite beam does not guarantee the globally highest-scoring K identifiers. The purpose of the constraint mechanism is to restrict the search frontier to eligible prefixes while preserving the score used to compare surviving candidates.

The SID mapping need not be assumed injective. If several entities share an identifier, generating that identifier establishes only that at least one associated entity is eligible. Entity-level output must resolve the SID against $E_{x,\tau}$, or apply an equivalent exact eligibility check. Model, tokenizer, SID mapping, and reverse lookup must also use compatible versions. Identifier validity is therefore necessary but not sufficient for entity-level correctness.

## 2.2 Prefix constraints

For a prefix h, define its legal next-token set as

$$A_{x,\tau}(h) = \{ v \in V : h{\cdot}v \text{ is a prefix of some } y \in S_{x,\tau} \}. \quad (3)$$

Only extensions in $A_{x,\tau}(h)$ enter the valid beam frontier. This condition removes a branch when no eligible identifier can complete it. It does not prove that every remaining branch is relevant to the user; relevance remains the responsibility of the retrieval model and downstream ranking.

The index represents one snapshot throughout a request. "Real-time" refers to constructing constraints from request-time supply, rather than periodically updating a single global catalog tree. It does not imply that the index changes continuously during decoding. Supply changes after τ require a freshness policy, revalidation, or a subsequent request.

## 2.3 Useful retrieval and compute

Let $U_x$ be the number of distinct eligible entities returned for request x after resolution and deduplication, and let $G_x$ be its measured accelerator time. Across a workload, one useful efficiency measure is

$$C_{useful} = \Sigma_x G_x / \Sigma_x U_x. \quad (4)$$

This aggregate includes resources spent on requests that return no useful result. Zero-result frequency must also be reported, and a workload with no useful outputs has no finite useful-output efficiency. CPU preparation, network traffic, and fallback costs should be reported alongside accelerator time.

A lower $C_{useful}$ is meaningful only at comparable retrieval quality and latency. At a fixed beam schedule, constraints may increase useful yield without reducing model FLOPs. A reduction in total serving cost requires sufficient useful-output gains, smaller decoding budgets at matched quality, fewer retries, or less fallback work to offset constraint overhead. We use these distinctions when evaluating the intended compute benefit.

## 3 System Architecture

### 3.1 Supply resolution and asset consistency

The online supply service combines request context with current serviceability to produce $E_{x,\tau}$. The SID mapping converts this entity set into $S_{x,\tau}$. The model prompt and the eligible identifiers then enter the inference service. After generation, reverse lookup produces business entities, and resolution, deduplication, and any required freshness checks precede downstream ranking.

This separation allows an existing retrieval model to react to supply changes without retraining for every weather or traffic event. It also defines the guarantee's boundary: an omitted eligible merchant cannot be recovered by the decoder, and a stale or incorrect supply set cannot be made current by a correct mask kernel. Eligibility completeness, snapshot age, and mapping compatibility are properties of the surrounding system.

An older global constraint tree remains useful for catalog validity and daily asset updates. It is not interchangeable with a per-request set. The former describes which identifiers exist in the deployed catalog; the latter describes which identifiers are eligible under a particular supply snapshot.

### 3.2 Request encoding and asynchronous compilation

The engine accepts three representations: a list of SID strings, a Base64-encoded array of packed 64-bit identifiers, and a Base64-encoded array of 16-bit offsets. In the offset representation, each three-token SID occupies six raw bytes, and the decoder adds the appropriate tokenizer base for each layer. Encoding width, layer bases, and mapping versions must agree with the deployed model.

The compact formats reduce the number of separately allocated JSON strings, but transport remains part of the request cost. The inference service submits parsing, sorting, and index construction to a CPU worker pool. This work can overlap model prefill once the eligible payload and prompt are available. The implemented overlap does not imply that upstream supply retrieval also overlaps prefill; that requires a separate cross-service protocol.

After sorting the SIDs, the builder identifies shared prefixes and emits their legal transitions. The request owns the resulting immutable index. On first use, the sampling path waits for compilation and transfers the index to the GPU. Subsequent steps reuse that index across all beams belonging to the request.

### 3.3 Physical CSR representation

The implementation specializes the prefix structure to L=3. A root mask records valid first tokens. The state reached by first token t has identifier t+1; two-token prefixes receive additional state identifiers. CSR row offsets identify contiguous arrays of legal token IDs and successor states. All beams share these arrays, while each active beam stores only its current constraint state.

This layout is sparse in its transitions but includes vocabulary-dependent allocation. The first-level state range is reserved across the full vocabulary, and the transition arrays include padding. Moreover, the GPU still materializes a dense beam-by-vocabulary mask. Sparse constraint storage therefore does not imply sparse model output projection or elimination of full-vocabulary mask memory.

### 3.4 Generation and output lifecycle

At each step, the engine computes model logits, obtains legal extensions from the current states, and selects the next beam frontier. The selected parent indices determine which histories, model states, and constraint states each child inherits. The selected token then advances the corresponding constraint state.

When fewer legal candidates exist than the configured beam width, only valid candidates should count toward the frontier and final output. The implementation propagates an effective beam count through sampling, dispatch, and stream updates. This count may increase at a later step when a surviving prefix has more children. Treating a temporary reduction to one beam as permanent greedy decoding would break that lifecycle.

## 4 Constrained Beam Execution

### 4.1 Preserving the retrieval score

Let $s_i$ be the accumulated log probability of parent beam i and $z_{iv}$ its logit for token v. Define $\lambda_i$ as the log-sum-exp over the full model vocabulary. A legal expansion receives

$$\text{score}(i,v) = s_i + z_{iv} - \lambda_i, \quad v \in A_{x,\tau}(h_i); \quad -\infty \text{ otherwise.} \quad (5)$$

If softmax is instead applied after masking, each parent's legal probability mass is renormalized to one. Let $m_i$ be the original probability mass assigned to its legal next tokens. The resulting score is shifted by $-\log m_i$. Because $m_i$ differs across parents, this operation can change their relative order.

For example, suppose two parent prefixes have probabilities 0.6 and 0.4. Each has one legal continuation, with original conditional probabilities 0.1 and 0.9. Their sequence probabilities are 0.06 and 0.36, so the second candidate should rank higher. Normalizing each singleton legal set produces scores proportional to 0.6 and 0.4, reversing the decision.

The implementation saves the full-vocabulary normalizer before applying the hard mask, subtracts it before cross-parent selection, and skips the beam kernel's internal log-softmax. Correcting scores after pruning cannot recover candidates already discarded under the wrong ranking. Each parent uses its own normalizer; the shared property is the scoring definition, not the numerical denominator.

This is an integration requirement rather than a new probability principle. STATIC also normalizes before masking [3]. The relevant issue in a serving engine is ensuring that preprocessing, fused sampling kernels, and later score handling collectively preserve that ordering. If additional processors transform logits, the normalizer must correspond to the logits that define the intended model score.

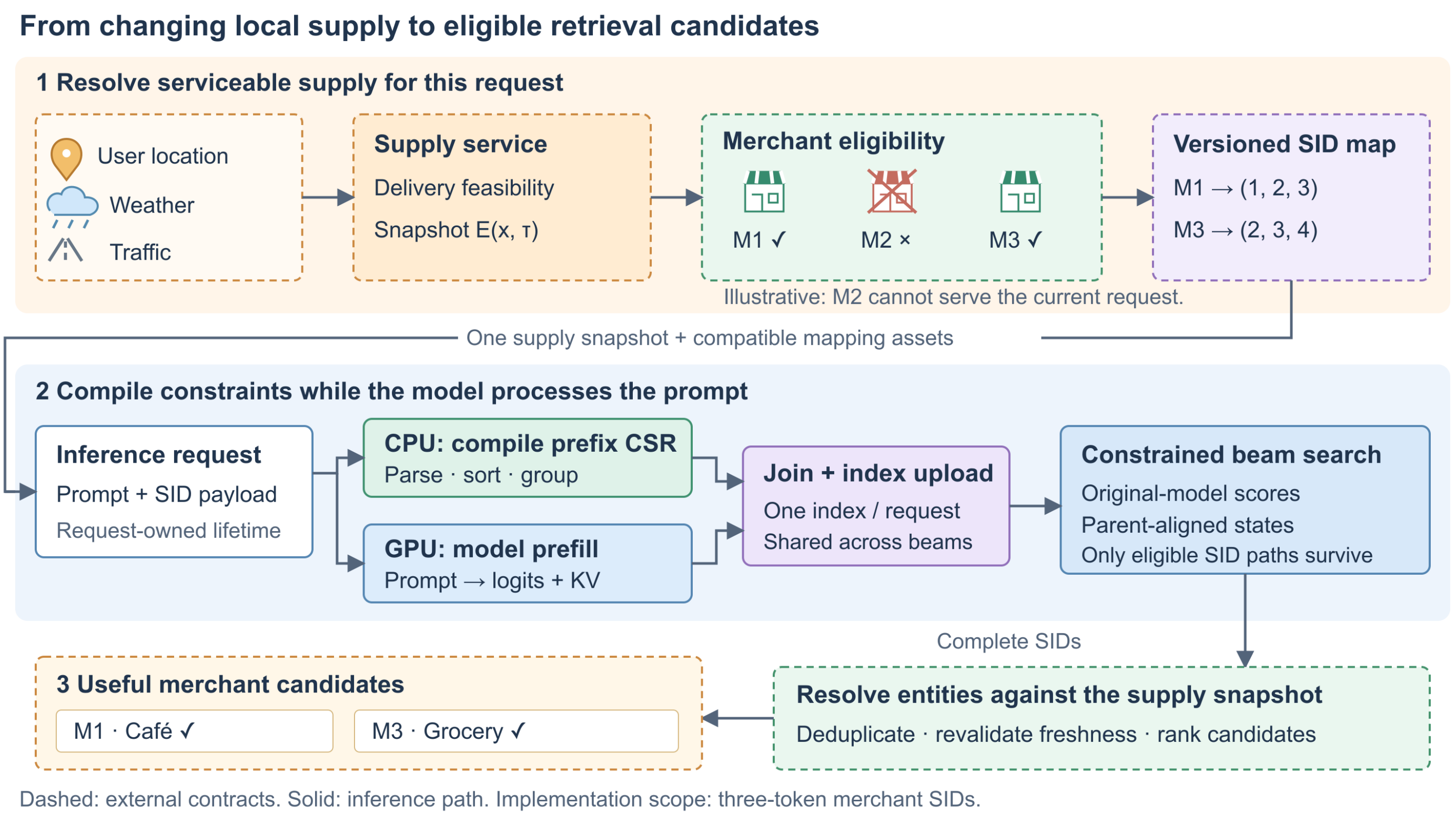

**Figure 1.** Request-time supply defines the feasible retrieval set. Location, weather, and traffic affect external serviceability decisions; the illustrative unavailable merchant M2 is excluded from the request constraint. CPU compilation overlaps model prefill, and beams share the uploaded index. Dashed boxes denote external contracts; solid boxes denote the inference implementation. The examined path generates three-token merchant SIDs.

## 4.2 Parent-aligned state propagation

Let $q_i$ be the constraint state of parent i, and let $(\pi_j, v_j)$ be a selected parent-token pair. The child's state is

$$q'_j = \delta(q_{\pi_j}, v_j). \quad (6)$$

The parent gather must occur before the transition. Applying $v_j$ to the old state at row j is incorrect whenever beam selection reorders parents. Multiple children may share a parent; each must inherit that parent's state and then apply its own selected token. The model's history and key-value cache must follow the same parent relation.

Algorithm 1 brings compilation, inference, selection, and state propagation into one request lifecycle. Algorithm 2 in Appendix C expands the constraint compiler used in its first line.

This algorithm assumes valid nonempty input, compatible mapping assets, and finite model normalizers. Line 14 specifies an empty-frontier exit in the logical contract; it is not a claim that every current runtime error path implements that behavior. Supply acquisition and entity resolution are external integration operations. The implementation maps lines 1-3 to asynchronous CSR initialization, lines 9-15 to the sampler, and lines 16-20 to stream/dispatcher updates. It reuses the GPU index but still performs the host-device state transfers discussed in Section 5.3.

## 4.3 Validity and score invariants

**Proposition 1 - Identifier validity.** Suppose the index exactly represents a valid nonempty set S of length-L identifiers; legal transitions alone enter the frontier; selected children inherit the correct parent states; and generation terminates at L. Every emitted complete identifier belongs to S.

**Proof.** The root represents the empty prefix, which prefixes every identifier in S. If a parent represents a prefix of S, its legal outgoing edges lead only to longer prefixes of S. Gathering the selected parent's state and then following its selected edge preserves this invariant for each child. Induction over L steps yields a complete length-L path in S. Invalid padding is excluded by assumption. [QED]

**Proposition 2 - Score preservation.** If each legal extension uses Equation (5), with the normalizer computed from the logits defining the model distribution, every surviving prefix retains its original-model accumulated log probability.

**Proof.** The empty prefix has log probability zero. Equation (5) adds the model's conditional log probability for the selected token to the parent's accumulated log probability. Parent-aligned gathering preserves the relevant prefix. Induction establishes the result. [QED]

These invariants do not establish exhaustive Top-K recovery, unbiased sampling, entity-level validity under ambiguous mappings, or freshness after the supply snapshot. The current physical representation also reuses state zero for terminal transitions and the root; the fixed-length stopping condition is therefore essential to the first proposition.

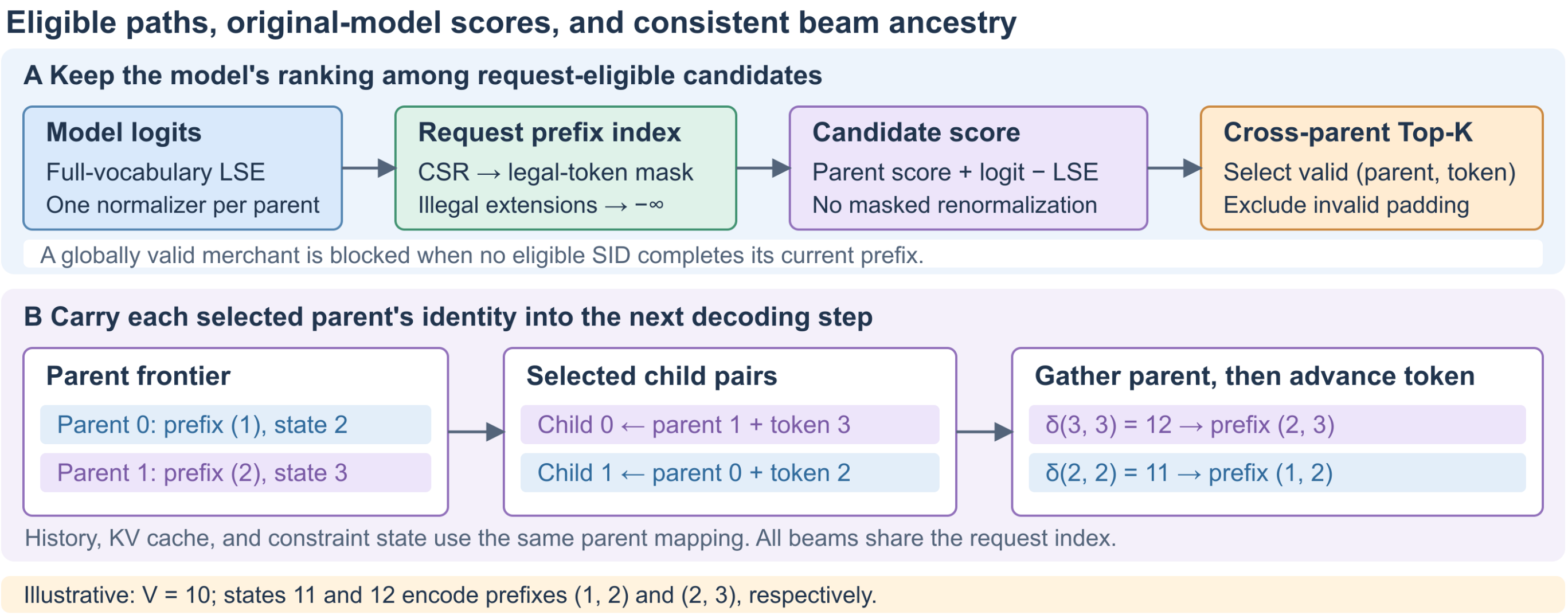

**Figure 2.** Score preservation and parent-aligned state propagation. Full-vocabulary normalization precedes hard masking and cross-parent selection. Colors track a parent's history, KV cache, and constraint state through reordering; each child then advances its selected token. The token and state IDs form an illustrative example, not a measured retrieval result.

---

**Algorithm 1. Core request-scoped supply-constrained retrieval.**

```
Require: prompt x; eligible entity snapshot E; compatible SID mapping φ; vocabulary size V
Require: valid nonempty S = {φ(e): e ∈ E}; L = 3; beam widths B[1..L]
Ensure: scored complete SIDs Y with identifiers(Y) ⊆ S; resolved entities R ⊆ E
1   f ← AsyncCPU(CompileRequestCSR(S, V))
2   (z, K) ← Prefill(x)
3   I ← UploadToGPU(Await(f))
4   H ← [empty prefix]; q ← [ROOT]; s ← [0]
5   for t = 1, ..., L do
6     if t > 1 then
7       (z, K) ← DecodeStep(K, LastTokens(H))
8     end if
9     λ[i] ← LogSumExp(z[i, :]) for every active parent i
10    M ← LegalNextTokenMask(I, q)
11    C ← {(i, v): M[i, v] = true and z[i, v] is finite}
12    a[i, v] ← s[i] + z[i, v] − λ[i] for (i, v) ∈ C
13    b ← min(B[t], |C|)
14    if b = 0 then return (empty, empty) end if
15    (π, v, s_next) ← TopKPairs(C, a, b)
16    K ← GatherParents(K, π)
17    H ← AppendTokens(GatherParents(H, π), v)
18    q_parent ← GatherParents(q, π)
19    q ← Transition(I, q_parent, v)
20    s ← s_next; effective_beams ← b
21  end for
22  Y ← {(H[j], s[j]): H[j] has length L and j is a valid beam}
23  R ← ResolveAndDeduplicate(Y, E, φ)
24  return (Y, R)
```

**Algorithm 1.** End-to-end logical flow. CPU compilation overlaps prefill (lines 1-2); the first-use join and upload precede constrained sampling (line 3). Original-model scores are preserved before cross-parent selection (lines 9-15), and the same parent mapping updates model and constraint states (lines 16-20). First-step logits come from prefill; later steps consume the previously selected tokens.

---

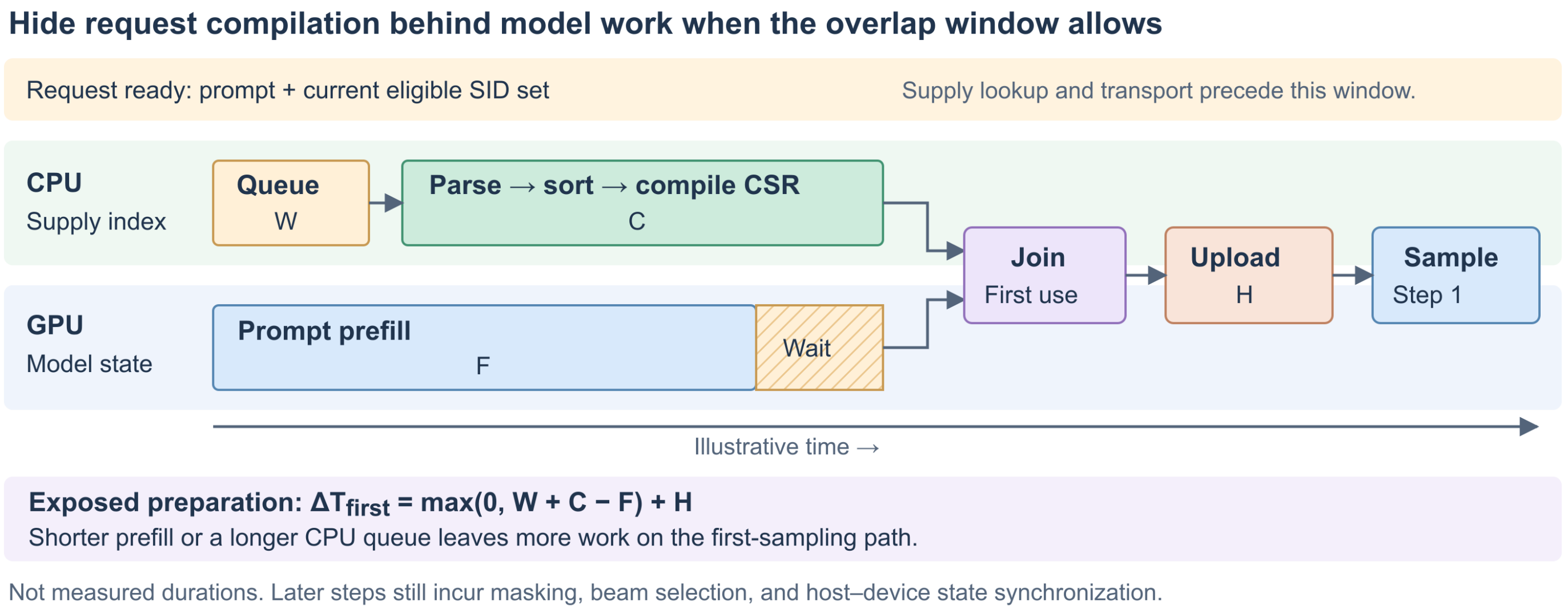

**Figure 3.** Dependencies on the first-sampling path. CPU compilation overlaps GPU prefill once the prompt and eligible SID payload are available. Short prefill or long compilation queues expose preparation latency; index upload remains on the critical path. Bar widths are illustrative. Later decoding steps still incur masking, selection, and host-device state synchronization.

# 5 Cost and Implementation Analysis

## 5.1 Exposed compilation latency

Let F be prefill time, W the CPU task's queueing delay, C its compilation time, and H its upload and associated first-use initialization time. If compilation and prefill begin from the same request-ready point and run independently, the first-sampling delay attributable to index preparation is

$$\Delta T_{first} = \max(0, W + C - F) + H. \quad (7)$$

Under these assumptions, overlap saves min(F,W+C) relative to serial preparation. Compilation is hidden when $W+C \le F$; upload is not. The formula excludes supply lookup, request transport, and work completed before submission to the worker pool. A late eligible payload or contention can increase exposed cost. This is particularly relevant for short prompts, which provide a smaller prefill window.

## 5.2 Index size and GPU work

For N fixed-length input SIDs, sorting requires O(N log N) comparisons and prefix grouping requires O(N) work, with L=3 treated as constant. Let V denote vocabulary cardinality, $P_2$ the number of distinct two-token prefixes, and U the number of distinct complete SIDs. The implementation's immutable GPU arrays occupy

$$I = 4(V+P_2+3) + 8(P_2+U+V) + V$$
$$= 13V + 12P_2 + 8U + 12 \text{ bytes}. \quad (8)$$

The terms account for 32-bit row offsets, 32-bit token and successor arrays including padding, and a one-byte root mask. At V=160,000, the vocabulary-dependent term alone is 2.08 MB in decimal units. For U=40,000, the complete expression is approximately 2.40-2.88 MB over $P_2 \in [1,40{,}000]$. These are layout-derived values, not measured peak memory; they exclude beam state, temporary construction buffers, logits, allocator overhead, and the dense mask.

For B active beams, the mask occupies B×V bytes. At a non-root state, the current kernel checks each vocabulary position against the state's outgoing row. If the maximum row degree is d, its worst-case membership work is O(BVd), followed by O(BV) mask application. The implementation does not yet eliminate dense vocabulary work or prune the model's output projection.

## 5.3 Communication and synchronization

A baseline that constructs and uploads a dense host mask every step transfers $V\Sigma_t B_t$ mask bytes for one-byte elements. The request-index approach instead uploads I once and incurs residual state traffic across steps. This reduces transfer volume only when the index plus residual traffic is smaller than the baseline's mask traffic; it is not a general end-to-end speedup guarantee.

The current implementation retains host involvement in the decoding loop. Output dispatch copies token data to the host, constraint updates may transfer tokens back to the device, updated states return to the host, and parent reordering prepares device states again. Immutable index residency should not be described as eliminating all host-device synchronization. This distinction matters for short decoding sequences where small transfers and synchronization can be a substantial fraction of added latency.

# 6 Historical Evaluation and Serving Evidence

## 6.1 Evaluation scope and stages

We examine three stages of archived evidence: T+1 daily-refresh merchant catalog constraints reported on 10 March 2026; incremental compiler and beam-execution measurements from April-May; and request-supplied constraints in May serving reports and captured-request replay [9-11]. These records precede the fixed June implementation snapshot in Appendix A. They describe system evolution, not a single version-matched benchmark. Hardware, deployed revision, and work-load metadata are incomplete for some records, and we identify those limits rather than combining measurements across stages.

The earlier T+1 stage refreshes the merchant constraint set daily. These constraints answer whether generated identifiers exist in that refreshed catalog. Request constraints additionally restrict generation to a supplied serviceable set. In serving tables, "no set" means no request-supplied SID set; it does not establish that every other catalog constraint was disabled. Reported entry counts and output beams are not assumed to be distinct eligible entities after resolution. We retain each source's timing labels; service RT, replay engine latency, and Whale-side latency are not assumed to use identical instrumentation.

Experimental location labels are anonymized. Northern city, Southeastern city, and Southwestern city denote cities in the corresponding regions of China; Southern district denotes a district in a city in southern China. The same labels are used consistently across the historical cases and serving reports.

### 6.2 Supply-update stages and post-generation filtering

Table 1 shows three archived T+1 daily-refresh cases [9] and a separately identified request-time row. In the Laoxiangji/Southern district case, both arms use beam schedule [50,150,300]; the measured initial count is 299, despite the configured maximum of 300. Total filtering decreases from 170 to 29, or 56.86% to 9.70% of generated merchants, a reduction of 47.16 percentage points. Surviving merchants increase from 129 to 270. Counts labeled hallucination-filtered fall from 120 to zero, and merchant-invalid counts from 24 to three; distance filtering remains 26 in both arms. The record does not formally define the first two categories. This is direct matched-beam case evidence for daily catalog constraints, not a request-level CSR effect or a population-wide estimate.

The other cases also reduce the source-labeled hallucination count to zero, while distance filtering persists or increases. This distinction motivates the request-time feasible set in Equation (1): a refreshed global catalog cannot encode every user's local serviceability. It does not, by itself, quantify the incremental benefit of request-scoped decoding.

Merchant gains do not guarantee product gains. In a separate store-product chain, both arms use merchant beams [50,100,200] and product beams [200,200,400]; surviving products in the Laoxiangji case increase from 33 to 44. Across the three product cases, survivors change from 33 to 40, 33 to 44, and 48 to 26, respectively. The last case worsens. Products did not yet have their own constraint mechanism, and the record notes downstream attachment and ranking adaptation issues. Online metric screenshots are reported separately in Appendix B; their comparison metadata are insufficient for a causal business-lift claim.

**Request-supplied constraints.** The subsequent implementation moves request-dependent supply eligibility into the set supplied with each request. Zero exclusions are expected for the eligibility predicates encoded in that set when downstream checks use the same snapshot, compatible SID mappings, and eligible-set-aware entity resolution. This property does not cover later supply changes or unrelated downstream filters.

### 6.3 Compilation and beam-execution diagnostics

The engineering record reports a three-run average of 7.70 ms for NumPy sorting and 8.63 ms for CSR construction on 33,766 triples [10]. A later C++ prototype reports approximately 3 ms and 1 ms, respectively (Table 2), with a Python/C++ sample comparison passing for 27,019 row-pointer entries and 60,137 packed token/state rows, including layout padding. The C++ timing paragraph does not restate all input and hardware conditions, so the arithmetic ratio is not a controlled speedup measurement.

Two observations explain why compiler and sampler integration matter. First, separate instrumented records report a 12.034 ms first sampling interval and an 11.579 ms first-use initialization path; these are not nested timings from the same request. Subsequent initialization checks in the latter record take 2 and 1 microseconds. The initialization measurement includes joining the compiler task and does not isolate CPU compilation. Flattened storage and packed-key sorting subsequently remove the visible wait in the recorded timeline, but the image does not quantify a zero residual. These observations support measuring the exposed term in Equation (7), rather than equating compilation duration with request overhead.

Second, strong masking creates many equal low logits and can make stable Top-K filtering expensive. One captured mask kernel takes 53.805 microseconds while a stable filtering kernel takes 3.892 ms. After changes to tie handling and execution paths, another radix kernel is recorded at 13.934 microseconds. These are different kernels, not a valid kernel-to-kernel speedup ratio. An April 24 comparison on a 0.4B model with 460 supplied entries reports enabled-minus-disabled latency differences of +0.22, -0.72, and +1.37 ms at beam widths 10, 100, and 200. Tie handling and overlap changed together; independent attribution and tie-order validation remain necessary.

Effective-beam bookkeeping also changes the validity of a performance baseline. A historical update condition failed when the effective width reached its configured maximum: schedule [50,150,300] produced only 150 outputs, and a shrinking schedule could fail. The corrected condition allowed the expected width to propagate. A comparison against the defective baseline measures unequal work; it cannot establish the constraint overhead for the corrected system.

### 6.4 Service and client latency after integration fixes

The May 18 record reports a 0.6 ms service-side RT increment for 30,000 supplied entries at each tested QPS (Table 3) [10]. The accompanying open-loop reports use 30 seconds of warmup and 180 seconds of steady-state load per level [11]. The no-set request is approximately 4.5 KB, with 351 input tokens, three output tokens, and beam schedule [50,150,300]. The supplied-set request is approximately 581 KB; its token configuration was not successfully parsed by the reporting script. The two reports were collected on different machines in the same service and at different times.

The service-side increment is about 2.3%, but the client increment at QPS 20 is 19.1 ms, or 69.2%; client p99 is 30.7 versus 55.9 ms. Thus, the small service RT difference cannot be advertised as the end-to-end overhead. The reports record successful completion without timeouts at these loads, but do not establish saturation throughput or a production tail-latency guarantee.

**Table 1.** Merchant filtering across supply-update stages. T+1 rows show constraint disabled / enabled; only the Laoxiangji case explicitly matches beam schedules [9]. † Request-time zeros are recalled by the lead author; original logs, workload, and sample size are unavailable. They are not matched measurements for the cases above. Dashes denote unreported values.

| Supply stage | Query / location | Initial count | Total filtered | Hallucination-filtered | Merchant-invalid | Distance-filtered |
|---|---|---|---|---|---|---|
| T+1 | McDonald's / Northern city | 300 | 149 / 45 | 94 / 0 | 12 / 0 | 43 / 45 |
| T+1 | Laoxiangji / Southern district | 299 | 170 / 29 | 120 / 0 | 24 / 3 | 26 / 26 |
| T+1 | Ham sausage / Southern district | 298 | 68 / 26 | 39 / 0 | 19 / 6 | 10 / 20 |
| Request-time | - | - | - | 0† | 0† | 0† |

**Table 2.** Reported preparation costs from separate development stages. The NumPy row averages three runs; the C++ record does not provide a matched measurement protocol. Neither row measures total request latency.

| Stage | Sort (ms) | Build (ms) | Sum (ms) |
|---|---|---|---|
| NumPy, 33,766 triples | 7.70 | 8.63 | 16.33 |
| Later C++ prototype | ~3 | ~1 | ~4 |

**Table 3.** May 18 records at the two reported timing boundaries. Service RT is transcribed from the engineering note; client means are from its attached load reports. These are different-host, different-window observations, not a same-host ablation. "No set" denotes no request-supplied SID set; the supplied-set arm contains 30,000 entries. Service RT's precise aggregation and instrumentation need confirmation.

| Target QPS | Service RT: no set (ms) | Service RT: supplied set (ms) | Client mean: no set (ms) | Client mean: supplied set (ms) |
|---|---|---|---|---|
| 10 | 26.1 | 26.7 | 28.2 | 47.1 |
| 15 | 25.8 | 26.4 | 28.1 | 47.3 |
| 20 | 25.4 | 26.0 | 27.6 | 46.7 |

The later report also corrects a monitoring error: the no-set report averaged utilization across an active and an idle machine, while the supplied-set report used a cross-machine maximum. For example, the corrected baseline GPU SM figure at QPS 20 is 15.82%, not 7.91%; the supplied-set report gives 17.52%. We omit resource columns from Table 3 because this correction and the different execution windows prevent direct causal attribution. Utilization alone would not establish per-request compute savings in any case.

## 6.5 Replay of captured supply requests

A subsequent report uses Base64 payloads captured from a test supply-service proxy path, representing the Southern district, Southeastern city, and Southwestern city scenarios [10, 11]. Table 4 summarizes the four groups, including a no-set baseline. Each group contains 20 displayed request records; all are marked PASS, but the report does not define that predicate, so it is not an entity-validity measurement.

For the Southern district scenario, the recomputed mean engine difference is approximately 0.03 ms, whereas the client difference is 1.52 ms. The Southwestern city and Southeastern city scenarios show lower latencies alongside fewer output beams. This is evidence that the implementation handles captured local-supply requests, but not a comparison at matched output quantity or retrieval quality. Client-minus-engine residuals vary with both input and output; they are not isolated network measurements. In particular, the 231-beam residual in the Southern district scenario exceeds the 300-beam baseline, so proportionality to output count alone is not supported.

We recomputed these statistics because the archived report's variability table is inconsistent with its request rows. Values labeled as engine standard deviations instead match the client sample standard deviations. Its claimed 47%-50% client variability is not supported by the displayed observations. Means and interpolated percentiles agree within the precision of the displayed data. Appendix B documents the evidence and reproducibility boundary.

## 6.6 What the evidence establishes

The records support increased surviving merchant yield in selected daily-constraint cases, including one explicitly matched-beam comparison, an implemented request-scoped pipeline, concrete compiler/sampler bottlenecks, and later serving behavior on captured supply payloads. They do not yet measure the incremental retrieval benefit of request-time constraints over daily constraints on a common workload, nor net compute savings at matched quality.

**Table 4.** May 25 captured-request replay, n=20 per group. Statistics are recomputed from displayed rows rounded to 0.01 ms; SD is sample standard deviation and p99 uses type-7 interpolation. Output beams are reported sequence counts, not deduplicated eligible entities. With 20 observations, p99 is near the sample maximum and is descriptive only. QPS, hardware, and the complete deployed configuration are not specified in this report.

| Scenario | Supply entries | Output beams | Client mean ± SD (ms) | Engine mean ± SD (ms) | Client p99 (ms) | Engine p99 (ms) |
|---|---|---|---|---|---|---|
| No set | 0 | 300 | 33.022 ± 1.958 | 23.307 ± 1.007 | 36.90 | 26.22 |
| Southwestern city | 2,836 | 92 | 26.771 ± 0.372 | 21.509 ± 0.131 | 27.63 | 21.80 |
| Southeastern city | 9,104 | 91 | 27.645 ± 0.389 | 21.655 ± 0.146 | 28.47 | 21.95 |
| Southern district | 26,419 | 231 | 34.543 ± 0.810 | 23.332 ± 0.151 | 36.58 | 23.60 |

A controlled follow-up should compare post-filtering, global catalog constraints, and request-supplied constraints with a common model, SID mapping, and timestamped supply snapshots. Fixed beam schedules measure useful-yield gains; matched useful recall or ranking quality measures resource savings. Both require entity resolution, deduplication, zero-result rates, CPU/GPU time, and end-to-end latency. Snapshot age and set change rate should be varied; geographic replay alone does not validate temporal adaptation or weather/traffic causality. Separate ablations of request encoding, compilation overlap, tie handling, and effective-beam propagation would identify which costs are hidden and which remain on the critical path.

## 7 Related Work

**Generative retrieval and semantic identifiers.** GENRE retrieves entities by autoregressively generating their names [1]. TIGER represents items with semantic identifiers and learns to predict the next item's identifier from interaction history [2]. These approaches establish the retrieval mechanism on which our system builds. Our focus is the serving contract between learned relevance and the merchant supply that can fulfill an individual request, with model training and SID construction treated as inputs.

**Catalog-constrained retrieval.** STATIC converts a prefix trie into accelerator-friendly CSR transitions [3]. Its design constructs the transition matrix offline, and its future-work discussion identifies dynamic sparse updates for real-time inventory changes. STATIC also normalizes logits before masking and propagates selected beam states. We therefore treat CSR storage, score-preserving masking, and parent-state propagation as established building blocks. Our system examines the additional preparation and synchronization costs incurred when an explicit supply set is compiled for each request. Comparisons must account for this preparation, rather than equating a per-step kernel measurement with full request latency.

**Personalized eligibility during generation.** GRACE already addresses request-dependent eligibility in real-time ads retrieval [4]. It periodically builds SID-prefix indexes containing targeting matchers, then evaluates request attributes during decoding. Prefix unions and Bloom matching can retain false positives, followed by exact ad-level filtering. Our design instead receives an explicit eligible SID set from an external supply service and compiles that set for the current request. This exchanges online compilation and transport costs for exact prefix membership relative to the supplied snapshot. It does not establish universal superiority: SID collisions still require entity resolution, and supply changes after the snapshot require a freshness policy.

**Dynamic structure and distributional objectives.** DISC combines dynamic importance sampling with GPU prefix verification to target asymptotically unbiased constrained sampling [5]. This objective differs from preserving original-model scores within finite beam search. XGrammar-2 supports changing output structures through dynamic dispatch, caching, and compilation mechanisms [6]. Dynamic constraints are therefore already an active research topic. Our specialization concerns short semantic identifiers and explicit local-supply snapshots under retrieval latency budgets.

**Positioning.** The systems distinction is between a prebuilt catalog index, a prebuilt rule index evaluated against request attributes, and a feasible set materialized and compiled for each request. We study the last setting. Its cost depends on supply freshness, set transmission, online preparation, and useful output yield as well as constraint execution. This distinction motivates an end-to-end comparison and avoids attributing novelty to request eligibility in general.

## 8 Limitations

The examined implementation supports a fixed three-token merchant SID path. Product-level supply introduces additional entity mappings and potentially larger eligible sets; its cost and correctness must be evaluated independently. The supply guarantee also depends on the accuracy and age of the upstream snapshot and on consistent model, tokenizer, and reverse-lookup assets.

The logical validity argument assumes valid nonempty input and correct exclusion of invalid slots. The current request boundary does not uniformly enforce this contract: empty constraint fields can select the older processing path, and malformed payload handling is permissive. Safe empty-supply and error behavior must therefore be validated separately from the valid-input algorithm.

Dense masks, full-vocabulary logits, linear row scans, and host-device state transfers remain in the current path. These limit claims about accelerator efficiency. The historical evidence spans selected daily-constraint cases, different-host load tests, and four groups of 20 replay requests. It does not quantify compute savings at matched retrieval quality, establish a causal online business lift, or isolate weather and traffic effects. Missing hardware, deployment, workload, and experiment metadata limit reproducibility and cross-stage attribution. The system provides a mechanism for avoiding supply-invalid search paths; its net production benefit depends on the operating point and requires the controlled evaluation described in Section 6.6.

## 9 Conclusion

Generative retrieval offers a promising direction for merchant and product discovery, but its output space must reflect the supply that each user can access. In local retail and food delivery, location, weather, and traffic change that space on a request-by-request basis. This work studies generative retrieval under dynamic supply constraints. Incorporating the eligible set into generation makes request-time feasibility part of the search process and can prevent beam capacity from being spent on infeasible candidates.

We describe a system that compiles eligible SIDs into a shared GPU prefix index, overlaps compilation with prefill, and preserves model scores and parent-aligned states during constrained beam execution. Its correctness and cost depend on the entire request lifecycle, including supply snapshots, transport, compilation, state synchronization, and entity resolution. Historical daily-constraint cases show increased surviving merchant yield, including a substantial gain in one explicitly matched-beam comparison, while leaving location filtering unresolved. Subsequent request-scoped engineering records show feasible online compilation and low reported service-side increments, but also expose client costs, effective-beam confounding, and measurement pitfalls. These findings motivate a joint evaluation of useful output and end-to-end resource use. Lower compute cost at matched retrieval quality remains an objective to establish, not an inference from engine latency alone.

## Appendix A Implementation Scope

The implementation analysis refers to the fixed snapshot in Reference 7. It covers request processor selection, parsing and CSR construction, asynchronous initialization, mask and transition kernels, score handling, and parent and effective-beam propagation. The code evidence index maps these components to source files and line numbers.

The full-vocabulary scoring integration was revised before cross-parent pruning in commit e929cb1; the later 6706ce2 change addresses token-history handling when effective beam width shrinks to one and subsequently expands. These version differences matter when matching measurements to behavior. The source snapshot is an implementation artifact, not a substitute for an executable correctness suite or a version-matched benchmark.

## Appendix B Additional Results and Measurement Provenance

### B.1 Online observations and product outcomes

Table 5 transcribes all five metrics in two archived online screenshots [9]. The first shows baseline group 437118 and a group without the daily-constraint treatment, 476821. The second shows daily-constraint group 486671 and its relative gap to a control, but does not display that control's ID again.

Observation windows, sample sizes, traffic allocation, and confidence intervals are absent. The two relative-gap columns therefore cannot be subtracted to estimate an incremental treatment effect. Displayed values are rounded, and screenshot color is not a substitute for a documented significance test.

The associated product-case counts retain mixed outcomes: McDonald's has 210 initial products and 177 versus 170 filtered, Laoxiangji has 232 and 199 versus 188 filtered, and ham sausage has 229 and 181 versus 203 filtered. These are merchant-treatment comparisons with no independent product constraint. They reinforce the need to evaluate downstream attachment and ranking, rather than assuming that more valid merchants imply better product or business outcomes.

### B.2 Earlier scale sweep

Table 6 preserves the earlier integration result used in the preceding manuscript [8]. The configuration is a 417-token prompt, beam schedule [50,150,300], and QPS 1. Whale-side latency excludes serialization and deserialization; the client observes a broader boundary. This earlier protocol/deployment should not be used as the cost of the later Base64 path.

**Table 5.** Archived daily-stage online observations. Each gap is the platform-reported relative difference to that image's control, not a direct daily-constraint versus no-treatment comparison. The second image's control identity and observation window remain unconfirmed. Conversion labels translate the source's visit-to-purchase metrics; their event-denominator definitions require confirmation.

| Metric | Control shown in first image | Without treatment | Reported gap | Daily-constraint group | Reported gap |
|---|---|---|---|---|---|
| Orders per user | 1.2977 | 1.3000 | +0.18% | 1.2994 | +0.13% |
| In-app order conversion | 71.13% | 71.23% | +0.14% | 71.24% | +0.15% |
| Recommendation conversion | 22.89% | 22.83% | -0.30% | 22.81% | -0.38% |
| Search conversion | 39.99% | 40.17% | +0.44% | 40.21% | +0.54% |
| Search orders per user | 0.3909 | 0.3926 | +0.44% | 0.3930 | +0.52% |

**Table 6.** Earlier request-size sweep, with no per-request records or uncertainty estimates supplied. The zero-entry row denotes no request-supplied set, not an empty-supply correctness test.

| Request entries | Client (ms) | Whale (ms) |
|---|---|---|
| 0, no supplied set | 26.5 | 26.0 |
| 10,000 | 34.0 | 27.0 |
| 20,000 | 39.6 | 27.7 |
| 30,000 | 46.0 | 29.0 |
| 40,000 | 54.0 | 29.8 |
| 50,000 | 61.0 | 31.5 |

At 50,000 entries, client and Whale increments are 34.5 and 5.5 ms. This motivates compact request handling but does not isolate an encoding speedup. The three geographic groups in Table 4 use different workloads and cannot be treated as a controlled before/after comparison with this sweep.

### B.3 Evidence and reproducibility boundary

The manuscript's historical measurement tables are drawn from archived internal notes, figures, and HTML reports [8-11]. The request-time row in Table 1 is attributed to the lead author's recollection provided on 5 November 2026; its original measurement records have not been recovered. The two source documents and three detailed reports were archived on 4 November 2026. The May 25 artifact contains 80 request rows with 0.01 ms display precision. We recompute means, sample standard deviations (denominator n-1), and quantiles using linear interpolation at (n-1)p. These are calculations on displayed measurements, not new inference runs or access to unrounded logs.

A companion author audit preserves source hashes, row-level CSV, recomputed statistics, and a standard-library parsing script. The internal reports and captured supply requests are not supplied as a public benchmark in this version. The public code reference identifies implementation scope but does not fix the hardware, model assets, or binary used in every historical test. Profiler attachments and original remote load-test logs have not been fully reanalyzed. A reproducible controlled evaluation still needs these assets, matched deployment configurations, input distributions, and documented validity checks.

## Appendix C Request Constraint Compilation

Algorithm 2 exposes the constraint preparation called by Algorithm 1. It describes the physical three-token representation, including its vocabulary-dependent state range and padding. Sorting and prefix deduplication compile the set; no model inference or supply feasibility prediction occurs in this procedure.

The physical implementation derives the same sorted prefix groups through adjacent comparisons; the named set P makes the representation explicit in pseudocode. Input decoding, token bounds, and mapping-version checks are preconditions rather than hidden operations in the CSR builder.

**Algorithm 2. Compile a three-token eligible set into a request CSR index.**

Require: valid nonempty token triples S; vocabulary size V

Ensure: root mask, row pointers, tokens, and successor states

```
T ← LexicographicSort(S); collapse duplicate triples
root_mask ← false[V]; edge_buckets ← empty
P ← sorted distinct two-token prefixes in T
Assign q(a, b) = V + 1 + rank_P(a, b), starting rank at 0
for each (a, b) ∈ P do
    root_mask[a] ← true
    AddEdge(edge_buckets[a + 1], token=b, next=q(a, b))
end for
for each distinct (a, b, c) ∈ T do
    AddEdge(edge_buckets[q(a, b)], token=c, next=0)
end for
Flatten buckets in state-ID order into tokens and next_states
Build indptr with V + |P| + 3 entries, including the padding row
Append V padding edges (token=V, next=0); close the padding
row
return (root_mask, indptr, tokens, next_states)
```

**Algorithm 2.** Root transitions are represented by root_mask and the implicit transition a → a+1. Non-root transitions occupy CSR rows. The terminal successor 0 is reused as the root representation, so Algorithm 1 must stop after exactly three SID tokens. Padding token V is outside the legal vocabulary [0,V), and is never an eligible continuation.